\documentclass{article}
\usepackage{amsmath}
\usepackage{latexsym}

\title{Nonperturbative Spontaneous Symmetry Breaking}
\author{Vladimir Dzhunushaliev
\thanks{
E-mail: dzhun@hotmail.kg}}
\begin{document}
\maketitle

\begin{center}
\textit{
Institut f\"ur Theoretische Physik,  
Freie Universit\"at Berlin, Arnimallee 14, D-14195, Berlin, Germany\\
and \\
Dept. Phys. and Microel. Engineer., Kyrgyz-Russian
Slavic University\\
Bishkek, Kievskaya Str. 44, 720000, Kyrgyz
Republic}
\end{center}

\begin{abstract}
A nonperturbative approach for  spontaneous symmetry breaking 
is proposed. It is based on some properties of \textit{interacting} field 
operators. As the consequences an additional terms like to 
$m^2 A^2$ appears in the initial Lagrangian. 
\end{abstract}

\section{Introduction}

One of the most astonishing results of quantum field theory is 
spontaneous symmetry breaking. One can say that it is the situation 
when something arises from the quantization. 
This means that on the quantum level one has something that was
not present on the classical level. Coleman and 
Weinberg \cite{coleman} write~: ``\dots higher-order effects may 
qualitatively change 
the character of a physical theory  \dots''. The main goal of the 
Coleman-Weinberg mechanism is to derive a Higgs potential from 
more fundamental principles, with as few arbitrary parameters as 
possible. In this mechanism the Higgs potential is induced by radiative 
corrections, rather than being inserted by hand. In this approach one can 
summarize over higher-loop graphs to induce an effective potential, 
which may then produce spontaneous symmetry breaking. 
Of coarse, it can be only in the theories with interactions 
(where we have these higher-loop graphs) : \textit{i.e.} a nonlinearity 
(in Lagrangian) can lead to the interesting consequences 
for quantized theory. 
It is not very surprisingly because on the classical level we have the 
same : very simple behavior of a classical linear theory can be changed 
on the very complicated and surprising behavior of a classical nonlinear 
theory. For example, in nonlinear theories we have monopoles, instantons, 
black holes, 
strange attractors and so on. On the quantum level we can expect once 
more amazing stuffs if we add some nonlinear terms in Lagrangian. 
Probably one of such manifestations of a nonlinearity is confinement 
in the QCD. 
\par 
The problem here is that we do not have detailed techniques for the 
nonperturbative calculations. Even on the perturbative level we do not 
sure that the result after the sum over all Feynman diagrams will the 
same as after the sum over a finite number of Feynman diagrams. 
Nevertheless, according to perturbative calculations we know that they change 
an initial Lagrangian so that an extra potential 
term arises in the Lagrangian (Coleman-Weinberg mechanism). 
\par 
In this paper we work out a nonpertutbative method which 
can be applied for strongly interacting fields and to show that extra terms 
(or nonperturbative spontaneous symmetry breaking) appear in an initial 
Lagrangian if the product of field operators have some 
properties concerning to the rearrangement of the brackets in 
a nonlinear potential term. 

\section{Nonlinear term in non-Abelian gauge theories}

Our basic attention here is devoted to the non-Abelian gauge field 
SU(2) (for the simplicity we will consider only this gauge group). 
The Lagrangian is 
\begin{equation}
\mathcal L = \frac{1}{4g^2}F^a_{\mu \nu} F^{a\mu \nu}
\label{sec1-10}
\end{equation}
where $F^a_{\mu \nu} = \partial_\mu A^a_\nu - \partial_\nu A^a_\mu + 
g \epsilon^{abc}A^b_\mu A^c_\nu$ is the field strength and $A^a_\mu$ 
is the gauge potential; $g$ is the coupling constant; 
$\epsilon^{abc}$ are the structural constants of the  
gauge group SU(2); $a=1,2,3$. In the quantum case we have the operators 
$\hat A^a_\mu$ and $\hat F^a_{\mu \nu} = \partial_\mu \hat A^a_\nu - 
\partial_\nu \hat A^a_\mu + g \epsilon^{abc}\hat A^b_\mu \hat A^c_\nu$. 
Let us underline that all operators considered here are the operators 
of \textit{interacting} fields in contrast with the perturbative 
techniques where these operators describe \textit{non-interacting} 
fields. 
\par 
Let us consider the nonlinear part of the field strength operator  
$\hat F^a_{\mu \nu}$ : 
$(\hat F_{nl})^a_{\mu \nu} = \epsilon ^{abc} \hat A^b_\mu \hat A^c_\nu$. 
At first we assume 
that this product do not have any singularity as it is the product 
of \textit{interacting} fields. Physically 
it means that there are situations in interacting 
field theories where these singularities do not occur 
({\it e.g.} for flux tubes in Abelian or non-Abelian theory 
quantities like the ``electric'' field inside the tube, 
$\langle E^a_z \rangle < \infty$, and energy density 
$\varepsilon (x) = \langle (E^a_z)^2 \rangle < \infty$ are 
nonsingular). Here we take as an assumption that such singularities 
do not occur. 
\par 
Thus we have such nonlinear term 
\begin{equation}
\begin{split}
\left (
  \left (
    \hat F_{nl}
  \right )^a_{\mu\nu}
\right )
\left (
  \left (
    \hat F_{nl}
  \right )^{a\mu\nu}
\right ) = 
\epsilon^{abc} \epsilon ^{ade} 
\left (
\hat A^b_\mu \hat A^c_\nu
\right )
\left (
\hat A^{d\mu} \hat A^{e\nu}
\right ) = &\\
\left (
\delta^{bd} \delta^{ce} - \delta^{be}\delta^{cd}
\right )
\left (
\hat A^b_\mu \hat A^c_\nu
\right )
\left (
\hat A^{d\mu} \hat A^{e\nu}
\right ) = & \\ 
\left (
\hat A^b_\mu \hat A^c_\nu
\right )
\left (
\hat A^{b\mu} \hat A^{c\nu}
\right ) - 
\left (
\hat A^b_\mu \hat A^c_\nu
\right )
\left (
\hat A^{c\mu} \hat A^{b\nu}
\right ) . 
\label{sec1-20}
\end{split}
\end{equation}
Our main assumption is that this nonlinear expression has 
\textit{some properties} that allows us to say that 
\begin{align}
\left (
\hat A^a_\mu \hat A^b_\nu
\right )
\left (
\hat A^{a\mu} \hat A^{b\nu}
\right ) - 
\left ( \left (
\hat A^a_\mu \hat A^b_\nu \right )
\hat A^{a\mu} \right ) \hat A^{b\nu} \neq 0 
\label{sec1-30}\\
\intertext{and}
\left (
\hat A^a_\mu \hat A^b_\nu
\right )
\left (
\hat A^{b\mu} \hat A^{a\nu}
\right ) - 
\left ( \left (
\hat A^a_\mu \hat A^b_\nu 
\right )
\hat A^{b\mu} \right )\hat A^{a\nu} \neq 0 
\label{sec1-40}
\end{align}
and these differences are not connected with the commutator  
$[\hat A^a_\mu , \hat A^b_\nu]$. The second term in Eq. \eqref{sec1-30} 
after some manipulations with 
brackets and permutations of $A^a_\mu$ can be led to 
$(\hat A^a_\mu \hat A^{a \mu})(\hat A^b_\nu \hat A^{b \nu})$. 
In this case this equation 
tell us that the square of nonlinear part $(F_{nl})^a_{\mu \nu}$ 
of the field strength is not equal to the square of the vector 
square $\hat A^4 = (\hat A^a_\mu \hat A^{a\mu})^2$, 
\textit{i.e.} the r.h.s. of Eq. \eqref{sec1-30} is nonzero. 
In some sense it is like to Cooper pairing in the superconductivity. 
The initial potential term for electrons is 
$\left (\hat\psi^+_\beta \left ( \hat\psi^+_\alpha \hat\psi_\alpha
\right ) \hat\psi_\beta \right )$ and after some manipulations 
we have 
$
\left (
\hat\psi_\alpha \hat\psi_\beta 
  \right ) 
\left (
\hat\psi^+_\gamma \hat\psi^+_\delta 
\right ) 
$
where every pair of brackets $(\cdots )(\cdots )$ has an independent 
physical meaning : each pair describes 
annihilation and creation of Cooper pair. We note that these pairs were 
not present in the initial Lagrangian. Another words, at first we had 
electrons and then Cooper pairs.  
\par 
The differences \eqref{sec1-30} \eqref{sec1-40} are connected only with 
the nonlinearity of the term $(\hat F_{nl})^2$. In order to calculate 
this difference we will consider at first a very simple example. 

\section{Simple example}

Let $\hat a$ and $\hat b$ are operators which have a nonassociative property
\begin{equation}
\left ( \hat a \hat b\right ) \left ( \hat a \hat b\right ) - 
\left ( \left (\hat a \hat b \right ) \hat a \right ) \hat b  
\neq 0 . 
\label{sec3-20}
\end{equation}
In order to calculate this commutations relationship we compare its 
with the ordinary commutators in a linear field theory 
\begin{equation}
\hat \phi (x) \hat \phi (y) - \hat \phi (y) \hat \phi (x) = 
-i \hbar D(x - y) 
\label{sec3-50}
\end{equation}
where $\hat \phi (x)$ is some field operator and $D(x - y)$ is some function. 
We see that at the l.h.s. we have the production of two 
operators and at the r.h.s. the number of operators is 
$(2-2)=0$. 
\par 
This procedure we would like to apply for the expression \eqref{sec3-20}. 
At the l.h.s. we have the production of four operators, consequently 
at the r.h.s. we should have the production of 
$(4-2)=2$ operators. Thus, we have two possibilities 
\begin{equation}
\left ( \hat a \hat b\right ) \left ( \hat a \hat b\right ) - 
\left ( \left ( \hat a \hat b \right ) \hat a \right ) \hat b = 
\lambda 
\begin{cases}
    0 , \; \text{if} \; \hat{b} = \hat{a} ; \\
    \lambda_1\hat{a}^2 + \lambda_2 \hat{b}^2, \; \text{if} \; 
    \hat{b} \neq \hat{a} . 
\end{cases} 
\label{sec3-60}
\end{equation}
The next simple example is for the vectors $A_\mu$. 
Let us determine the following relation 
\begin{equation}
  \left( 
  \hat{A}_\mu \hat{A}_\nu
  \right)
  \left( 
  \hat{A}^\mu \hat{A}^\nu
  \right ) - 
  \left( \left( \hat{A}_\mu \hat{A}_\nu 
  \right) \hat{A}^\mu \right ) \hat{A}^\nu = 
  \lambda 
  \begin{cases}
    0 , \; \text{if} \; \mu = \nu \\
    \hat{A}_\nu \hat{A}^\nu , \; \text{if} \; 
    \mu \neq \nu .
  \end{cases}    
\label{sec3-70}
\end{equation}
Here and up to end we will write $\sum$ if there is the sum, consequently 
in this equation \eqref{sec3-70} we have not any sum over repeating 
indices. Now we summarize over $\nu$ 
\begin{equation}
\begin{split}
  \sum^3_{\nu = 0} \left( 
  \hat{A}_\mu \hat{A}_\nu
  \right)
  \left( 
  \hat{A}^\mu \hat{A}^\nu
  \right ) - 
  \sum^3_{\nu = 0} \left( \left( \hat{A}_\mu \hat{A}_\nu 
  \right) \hat{A}^\mu \right ) \hat{A}^\nu = 
  \lambda \sum_{\nu \neq \mu} \hat A_\nu \hat A^\nu = \\ 
  \lambda 
  \left(
  \sum^3_{\nu =0 } \hat A_\nu \hat A^\nu - 
  \hat A_\mu \hat A^\mu 
  \right). 
\label{sec3-80}
\end{split}
\end{equation}
After the sum over $\mu$ we have 
\begin{equation}
  \sum^3_{\mu , \nu = 0} \left( 
  \hat{A}_\mu \hat{A}_\nu
  \right)
  \left( 
  \hat{A}^\mu \hat{A}^\nu
  \right ) - 
  \sum^3_{\nu = 0} \left( \left( \hat{A}_\mu \hat{A}_\nu 
  \right) \hat{A}^\mu \right ) \hat{A}^\nu = 
  3 \lambda \sum_{\nu = 0} \hat A_\nu \hat A^\nu 
\label{sec3-90}
\end{equation}
We would like to underline again that the differences \eqref{sec3-60} 
\eqref{sec3-90} 
\textit{are not connected with the commutator} $[\hat a , \hat b]$ 
which can be zero. This is only a manifestation of the nonlinearity 
of the corresponding expression 
$( \hat a \hat b ) ( \hat a \hat b )$. 

\section{Non-Abelian case}

Let us come back to more realistic case : non-Abelian gauge 
theories. We suppose that for the first term of the r.h.s. of 
Eq. \eqref{sec1-20} we have such equation 
\begin{equation}
\begin{split}
\left (
\hat A^a_\mu \hat A^b_\nu
\right )
\left (
\hat A^{a\mu} \hat A^{b\nu}
\right ) - 
\left (
  \left (
  \hat A^a_\mu \hat A^b_\nu
  \right ) \hat A^{a\mu}
\right ) \hat A^{b\nu} = \\
\lambda 
  \begin{cases}
    0 , \; \text{if} \; b =a , \nu = \mu ; \\
    \hat{A}^a_\nu \hat{A}^{a \nu} , \; \text{if} \; b = a , \nu \neq \mu ; \\
    \hat{A}^b_\mu \hat{A}^{b \mu} , \; \text{if} \; b \neq a , \nu = \mu ; \\
    \hat{A}^b_\nu \hat{A}^{b \nu} , \; \text{if} \; 
    b \neq a , \nu \neq \mu .
  \end{cases}   
\label{sec4-10b}
\end{split}
\end{equation}
Let us remind that here we have not the sum over repeating indices. 
Using the same calculations as in the previous section we have after 
summarizing over $b, \nu$ 
\begin{equation}
\begin{split}
\sum_{b, \nu} 
  \left (
  \hat A^a_\mu \hat A^b_\nu
  \right )
  \left (
  \hat A^{a\mu} \hat A^{b\nu}
  \right ) - 
  \sum_{b, \nu} 
  \left( \left( \hat A^a_\mu \hat A^b_\nu \right ) 
  \hat A^{a\mu} \right ) \hat A^{b\nu} = & \\
\lambda 
\left(
    \sum_{\nu \neq \mu} \hat{A}^a_\nu \hat{A}^{a \nu} + 
    \sum_{b \neq a} \hat{A}^b_\mu \hat{A}^{b \mu} + 
    \sum_{b \neq a , \nu \neq \mu} \hat{A}^b_\nu \hat{A}^{b\nu} 
\right) = & \\
\lambda 
\left(
\sum_{b , \nu} \hat{A}^b_\nu \hat{A}^{b\nu} - 
\hat{A^a_\mu} \hat{A}^{a \mu} 
\right) & . 
\label{sec4-10c}
\end{split}
\end{equation}
Finally, after summarizing over $a, \mu$ we have 
\begin{equation}
\begin{split}
\sum_{a, \mu} \sum_{b, \nu} 
  \left (
  \hat A^a_\mu \hat A^b_\nu
  \right )
  \left (
  \hat A^{a\mu} \hat A^{b\nu}
  \right ) = & \\
\sum_{a, \mu} \sum_{b, \nu} 
  \left( \left( \hat A^a_\mu \hat A^b_\nu \right ) 
  \hat A^{a\mu} \right ) \hat A^{b\nu} + 
11 \lambda   
\sum_{b , \nu} \hat{A}^b_\nu \hat{A}^{b\nu} . 
\label{sec4-10d}
\end{split}
\end{equation}
For the next term 
$(\hat A^a_\mu \hat A^b_\nu )
(\hat A^{b\mu} \hat A^{a\nu} ) 
$
in Eq. \eqref{sec1-20} we have some modifications. We have to calculate 
the following difference 
\begin{equation}
  \sum_{a, \mu} \sum_{b, \nu}
  \left(
  \hat{A}^a_\mu \hat{A}^b_\nu
  \right) 
  \left( 
  \hat{A}^{b \mu} \hat{A}^{a \nu} 
  \right) - 
  \sum_{a, \mu} \sum_{b, \nu} 
  \left( \left( \hat{A}^a_\mu \hat{A}^b_\nu 
  \right) \hat{A}^{b \mu} \right) \hat{A}^{a \nu} = ? 
\label{sec4-15}
\end{equation}
At first we will consider the next cases with different relations 
between $b,a$ and $\nu , \mu$.
\par
\noindent
If $b = a$ and $\nu = \mu$ 
\begin{equation}
\left (
\hat A^a_\mu \hat A^a_\mu
\right )
\left (
\hat A^{a\mu} \hat A^{a\mu}
\right ) - 
\left (
  \left (
  \hat A^a_\mu \hat A^a_\mu
  \right ) \hat A^{a\mu}
\right ) \hat A^{a\mu} = 0 .  
\label{sec4-40a}
\end{equation}
If $b = a$ and $\nu \neq \mu$ 
\begin{equation}
\left (
\hat A^a_\mu \hat A^a_\nu
\right )
\left (
\hat A^{a\mu} \hat A^{a\nu}
\right ) - 
\left (
  \left (
  \hat A^a_\mu \hat A^a_\nu
  \right ) \hat A^{a\mu}
\right ) \hat A^{a\nu} = 
\lambda \hat{A}^a_\nu \hat{A^{a \nu}} . 
\label{sec4-40b}
\end{equation}
If $b \neq a$ and $\nu = \mu$ 
\begin{equation}
\left (
\hat A^a_\mu \hat A^b_\mu
\right )
\left (
\hat A^{b\mu} \hat A^{a\mu}
\right ) - 
\left (
  \left (
  \hat A^a_\mu \hat A^b_\mu
  \right ) \hat A^{b\mu}
\right ) \hat A^{a\mu} = 
\lambda \hat{A}^b_\mu \hat{A^{b \mu}} . 
\label{sec4-40c}
\end{equation}
If $b \neq a$ and $\nu \neq \mu$ 
\begin{equation}
\left (
\hat A^a_\mu \hat A^b_\nu
\right )
\left (
\hat A^{b\mu} \hat A^{a\nu}
\right ) - 
\left (
  \left (
  \hat A^a_\mu \hat A^b_\nu
  \right ) \hat A^{b\mu}
\right ) \hat A^{a\nu} = 0 .
\label{sec4-40d}
\end{equation}
The relation \eqref{sec4-40d} demands more careful discussion. 
In fact we assume here that the difference \eqref{sec4-15} is nonzero 
only if there is at least two identical factors 
(as in Eq's \eqref{sec4-40a}-\eqref{sec4-40c}). Let us consider the 
situation with two identical pairs 
$(\hat A^a_\mu \hat A^b_\nu)(\hat A^{a \mu} \hat A^{b \nu})$. 
After some permutations of brackets and factors we can derive 
$(\hat A^a_\mu \hat A^{a\mu})(\hat A^b_\nu \hat A^{b \nu})$. 
If we average some quantum state $| Q \rangle$ we will have 
\begin{equation}
\begin{split}
  \left\langle Q | 
  \left( \hat A^a_\mu \hat A^b_\nu \right) 
  \left( \hat A^{a\mu} \hat A^{b\nu} \right)
  | Q \right\rangle - 
  \left\langle Q | 
  \left( \hat A^a_\mu \hat A^{a\mu} \right) 
  \left( \hat A^b_\nu \hat A^{b\nu} \right)
  | Q \right\rangle = \\
  \lambda 
  \left\langle Q | 
  \left( \hat A^b_\nu \hat A^{b\nu} \right)
  | Q \right\rangle + 
  \bigl( something \bigl)
\label{sec4-45}
\end{split}
\end{equation}
here $(something)$ is connected with commutators and permutations of 
brackets for three factors (for example, 
$(\hat A^a_\mu \hat A^b_\nu) \hat A^{a \mu} \rightarrow 
\hat A^a_\mu (\hat A^b_\nu \hat A^{a \mu})$ 
and so on). It means that the difference \eqref{sec4-45} can be nonzero 
only if a few factors are identical but if all factors are different then 
the first term on the r.h.s. of Eq. \eqref{sec4-45} is absent. Physically, 
one can say that the quantum correlation between four different components 
of the vector potential is the same for any combination of brackets 
(with an accuracy of the second term in Eq. \eqref{sec4-45}). 
\par 
In view of Eq's \eqref{sec4-40a}-\eqref{sec4-40c} the sum over 
$b, \nu$ in Eq. \eqref{sec4-15} give us 
\begin{equation}
\begin{split}
  \sum_{b, \nu} 
  \left(
  \hat{A}^a_\mu \hat{A}^b_\nu
  \right) 
  \left( 
  \hat{A}^{b \mu} \hat{A}^{a \nu} 
  \right) - 
  \sum_{b, \nu} 
  \left( \left( \hat{A}^a_\mu \hat{A}^b_\nu 
  \right) \hat{A}^{b \mu} \right) \hat{A}^{a \nu} = & \\
  \lambda 
  \left(
  \sum_{\nu \neq \mu} \hat{A}^a_\nu \hat{A}^{a\nu} + 
  \sum_{b \neq a} \hat{A}^b_\mu \hat{A}^{b\mu}
  \right) = 
  \lambda 
  \left(
  \sum_\nu \hat{A}^a_\nu \hat{A}^{a\nu} + 
  \sum_b \hat{A}^b_\mu \hat{A}^{b\mu} - 
  2 \hat{A}^a_\mu \hat{A}^{a\mu}
  \right) . 
\label{sec4-50}
\end{split}
\end{equation}
Now we can summarize over $a, \mu$ and Eq. \eqref{sec4-15} 
has the following form 
\begin{equation}
\begin{split}
  \sum_{a, \mu} \sum_{b, \nu}
  \left(
  \hat{A}^a_\mu \hat{A}^b_\nu
  \right) 
  \left( 
  \hat{A}^{b \mu} \hat{A}^{a \nu} 
  \right) - 
  \sum_{a, \mu} \sum_{b, \nu} 
  \left( \left( \hat{A}^a_\mu \hat{A}^b_\nu 
  \right) \hat{A}^{b \mu} \right) \hat{A}^{a \nu} = & \\
  \lambda 
  \left(
  4 \sum_{a,\nu} \hat{A}^a_\nu \hat{A}^{a\nu} + 
  3 \sum_{b, \mu} \hat{A}^b_\mu \hat{A}^{b \mu} - 
  2 \sum_{a, \mu} \hat{A}^a_\mu \hat{A}^{a \mu}
  \right) = 
  5 \lambda \sum_{a, \mu} \hat{A}^a_\mu \hat{A}^{a \mu} . 
\label{sec4-60}
\end{split}
\end{equation}
Consequently 
\begin{equation}
\begin{split}
\left (
  \left (
    \hat F_{nl}
  \right )^a_{\mu\nu}
\right )
\left (
  \left (
    \hat F_{nl}
  \right )^{a\mu\nu}
\right ) = 
\left (
\hat A^a_\mu \hat A^b_\nu
\right )
\left (
\hat A^{a\mu} \hat A^{b\nu}
\right ) - 
\left (
\hat A^a_\mu \hat A^b_\nu
\right )
\left (
\hat A^{b\mu} \hat A^{a\nu}
\right ) = & \\
\left( \left( \hat A^a_\mu \hat A^b_\nu \right ) 
  \hat A^{a\mu} \right ) \hat A^{b\nu} - 
\left( \left( \hat{A}^a_\mu \hat{A}^b_\nu 
  \right) \hat{A}^{b \mu} \right) \hat{A}^{a \nu} + 
6 \lambda \hat{A}^a_\mu \hat{A}^{a \mu} 
\label{sec4-70}
\end{split}
\end{equation}
here we again restore the ordinary rule for the sum over repeating 
indices. It is evidently that our extra term 
$(6 \lambda A^a_\mu A^{a\mu})$ 
breaks the initial gauge symmetry of given Lagrangian \eqref{sec1-10} 
and consequently it is similar to Coleman-Weinberg symmetry 
breaking but on the nonperturbative level. 
\par 
It is necessary to note that all these calculations 
\eqref{sec4-10b}-\eqref{sec4-70} was done for the fields 
in one point $(x)$. 

\section{Some explanations}

Now we would like to explain why we need with equation 
\eqref{sec4-70}. Let us assume that we have a quantum 
state $| Q \rangle$ and the action of the field operator 
$\hat A^a_\mu$ on this state is 
$|\Phi_1 \rangle = \hat A^a_\mu |Q \rangle$. For the nonlinear term the 
problem is that we do not know the action of this term by the 
direct way 
\begin{equation}
\left (
\hat A^a_\mu \hat A^b_\nu
\right )
\left (
\hat A^{a\mu} \hat A^{b\nu} 
\right ) | Q \rangle = ?
\label{sec5-10}
\end{equation}
because we do not know the action of the operator 
$(\hat A^{a\mu} \hat A^{b\nu})$ on the quantum state $|Q \rangle$. 
In order to determine the action \eqref{sec5-10} we need for the 
following transformations 
\begin{eqnarray}
\left (
\hat A^a_\mu \hat A^b_\nu
\right )
\left (
\hat A^{a\mu} \hat A^{b\nu} 
\right ) | Q \rangle \rightarrow 
\left (
  \left (
  \hat A^a_\mu \hat A^b_\nu
  \right )\hat A^{a\mu} 
\right ) 
\left (
\hat A^{b\nu}  | Q \rangle 
\right ) = && 
\nonumber \\
\left (
\hat A^a_\mu \hat A^b_\nu
\right )
\left (
\hat A^{a\mu} | \Phi_1 \rangle 
\right ) = 
\hat A^a_\mu 
\left (
\hat A^b_\nu | \Phi_2 \rangle 
\right ) = 
\hat A^a_\mu | \Phi_3 \rangle = 
| \Phi_4 \rangle && 
\label{sec5-20}
\end{eqnarray}
where $| \Phi_1 \rangle = \hat A^a_\mu | Q \rangle$ and so on 
for every $| \Phi_i \rangle$, $i=2,3,4$. 

\section{Discussions and conclusions}

In this paper we have shown that a nonlinear potential in 
quantum non-Abelian gauge theory can lead to the appearance 
of some extra terms in Lagrangian. In Ref. \cite{dzhsin} 
it is shown that Ginzburg - Landau equation can be derived from a pure 
non-Abelian gauge theory. There was assumed that the quantization 
of such theory leads to the appearance of $m^2 A^2$-like 
term in the initial Lagrangian. In this paper we suggested a 
nonperturbative mechanism for this phenomenon. 
\par 
A possibility that the operators of quantum fields with a nonlinear 
potential can have nonassociative properties was investigated 
in Ref. \cite{dzhsin2} . In this paper we continue this approach 
but there is one very essential disctinction : we assume that 
the remainders in Eq's \eqref{sec1-30} \eqref{sec1-40} are not 
a numerical function 
but is an operator which is the product of fields operators. 
\par
In Ref's \cite{dzhsin1} it is shown that a Meissner-like effect 
in non-Abelian gauge theories arises. There was applied 
a nonperturbative quantization techniques based on the 
Heisenberg's approach to a nonlinear spinor field. 
Heisenberg's idea is to  use field operator 
equations for receiving an infinite 
equations set for all Green's functions. In this paper (remaining in 
the frame of Heisenberg's approach) we show that this nonperturbative 
mechanism leads to spontaneously symmetry breaking. 
\par 
Finally, we would like to say that probably nonperturbative 
quantum field theory with a strong interaction 
will have very unusual and unexpected properties 
in contrast with quantum field theory with a small coupling constant. 
This difference is similar with the difference between analytical 
and differentiable (but nonanalytical) functions. The first case 
is similar to a quantum field theory with a small coupling constant, 
where we can expand a function in Taylor-series which for 
the quantum theory is Feynman diagrams. In the second case a 
function can not be expanded in any Taylor-series and respectively Feynman 
techniques can not be applied for such kind of quantum field theories. 
Probably, the QCD belong to the last case that leads to the fact that 
confinement can not be explained on the language of Feynman diagrams. 
In this case one can say \cite{hansson} that ''fields are primary 
to particles''. 
\par 
Our opinion is that in this case we can use the Heisenberg's 
nonperturbative quantization method \cite{heis}. 
Something like this takes place in the classical case~: 
the nonlinear classical theories have such nonperturbative phenomena 
as~: self-ogranisation, monopoles, instantons, black holes, strange attractors 
and so on which are absent in linear theories and they can not be 
derived by the perturbative way. 

\section{Acknowledgments}
I am very grateful for the DAAD for the financial support and 
Prof. Hagen Kleinert for the hospitality in his research group. 
Partially this work was supported by Alexander von 
Humboldt Foundation.


\begin{thebibliography}{99}


\bibitem{coleman}
S. Coleman and E. Weinberg, Phys. Rev., \textbf{D7}, 1888 (1973).

\bibitem{dzhsin}
V. Dzhunushaliev and D. Singleton, ``Ginzburg - Landau equation 
from SU(2) gauge field theory'', in preparation. 

\bibitem{dzhsin2}
V. Dzhunushaliev, Theor. Math. Phys., \textbf{100}, 1082 (1994); 
V. Dzhunushaliev, ``Nonperturbative operator 
quantization of strongly nonlinear fields'', hep-th/0103172; 
V. Dzhunushaliev and D. Singleton, Nuovo Cimento, \textbf{117B}, 
137 (2002). 

\bibitem{dzhsin1}
V. Dzhunushaliev and D. Singleton, Phys. Rev., \textbf{D65}, 
125007 (2002).

\bibitem{hansson}
J. Hansson, ``A simple explanation of the non-appearnace of physical 
gluons ans quarks'', hep-ph/0208137, to appear in Can. J. Phys. 

\bibitem{heis}
W. Heisenberg, \textit{Introduction to the unified field theory of
elementary particles.}, Max - Planck - Institut f\"ur Physik und
Astrophysik, Interscience Publishers London, New York, Sydney,
1966; W. Heisenberg, Nachr. Akad. Wiss. G\"ottingen, N8,
111(1953); W. Heisenberg, Zs. Naturforsch., \textbf{9a},
292(1954); W. Heisenberg, F. Kortel und H. M\"utter, Zs.
Naturforsch., \textbf{10a}, 425(1955); W. Heisenberg, Zs. f\"ur
Phys., \textbf{144}, 1(1956); P. Askali and W. Heisenberg, Zs.
Naturforsch., \textbf{12a}, 177(1957); W. Heisenberg, Nucl. Phys.,
\textbf{4}, 532(1957); W. Heisenberg, Rev. Mod. Phys., \textbf{29},
269(1957).

\end{thebibliography}
\end{document}